\documentclass[twocolumn,amsmath,amssymb,nofootinbib,floatfix,prl,superscriptaddress,showpacs]{revtex4}

\usepackage{graphicx}
\usepackage{bm, bbm}      
\usepackage{epstopdf}

\begin{document}

\title{Electron fractionalization in two-dimensional graphenelike structures}

\author{Chang-Yu Hou}
\author{Claudio Chamon}
\affiliation{
Physics Department, Boston University,
590 Commonwealth Ave., Boston, MA 02215, USA
            }

\author{Christopher Mudry}
\affiliation{
Condensed matter theory group,
Paul Scherrer Institut,
CH-5232 Villigen PSI , Switzerland
            }
\date{\today}

\begin{abstract}

Electron fractionalization is intimately related to topology. In
one-dimensional systems, fractionally charged states exist at domain
walls between degenerate vacua. In two-dimensional systems, 
fractionalization exists in quantum Hall fluids, where time-reversal symmetry is
broken by a large external magnetic field. Recently, there has been a
tremendous effort in the search for examples of fractionalization in
two-dimensional systems with time-reversal symmetry. In this letter, we
show that fractionally charged topological excitations exist on
graphenelike structures, where quasiparticles are described by two
flavors of Dirac fermions and time-reversal symmetry is respected. The
topological zero-modes are mathematically similar to fractional
vortices in $p$-wave superconductors. They correspond to a twist in the
phase in the mass of the Dirac fermions, akin to cosmic strings in
particle physics.
\end{abstract}


\maketitle
In low dimensional systems, the excitation spectrum sometimes contains
quasiparticles with fractionalized quantum numbers. A famous
example of fractionalization was obtained in one-dimension (1D)
by Jackiw
and Rebbi~\cite{Jackiw-Rebbi} and by Su, Schrieffer and
Hegger~\cite{Su1979}. They showed the existence of charge
$e/2$ states, with polyacetelene as a physical realization of such
phenomena. In these systems, a charge density wave develops and the
ground state is two-fold degenerate. The fractionalized states
correspond to mid-gap or zero-mode solutions that are sustained at the
domain wall (a soliton) interpolating between the two-degenerate
vacua.

The fractional quantum Hall effect provides an example of
fractionalization in two-dimensions (2D). Not only do the Laughlin
quasiparticles have fractional charge~\cite{Laughlin}, but they also
have fractional (anyon)
statistics~\cite{Halperin,Arovas}. Time-reversal symmetry (TRS) is broken
due to the strong magnetic field, leaving as an outstanding problem
the search for systems where fractionalization is realized without the
breaking of TRS. The motivation for such a quest 
stems from speculations that fractionalization may play a role in the 
mechanism for high-temperature superconductivity%
~\cite{Anderson1987,Kivelson87,Laughlin1988}. 
Progress has been made on finding model systems, such as dimer models, in which
monomers defects act as fractionalized (and deconfined, in the case of
the triangular lattice) excitations. 

In this letter, we present a mechanism to fractionalize the electron
in graphenelike systems that leaves TRS unbroken. The
excitation spectrum of honeycomb lattices, which have been known
theoretically for a few decades to be described by Dirac
fermions~\cite{Wallace1947,Semenoff1984}, is now the subject of many recent
studies since single and few atomic-layer graphite samples have been
realized experimentally~\cite{Novoselov}. Quantum number fractionalization
is intimately related to topology and here we find that a twist or a
vortex in an order parameter for a mass gap gives rise to a single
mid-gap state at zero energy. Such twist in the mass of the
Dirac fermions in graphenelike structure is the analogous in 2+1 space-time
dimensions of a cosmic string in 3+1-dimensions%
~\cite{Witten:1984eb}. 

The zero-modes we find are, in their mathematical structure, similar
to those found in $p$-wave superconductors by Read and
Green~\cite{Read-Green} and to those found by Jackiw and
Rossi~\cite{Jackiw-Rossi} when demonstrating topological excitations
and suggesting an index theorem in 2D
(see also Ref.~\onlinecite{Cugliandolo89}). 
In Refs.~\onlinecite{Read-Green,Jackiw-Rossi},
and~\onlinecite{Cugliandolo89},
a twist in the phase of 
a superconducting order parameter
and a charge $2e$ Higgs boson,
respectively,
were considered and thus the electric charge assigned
to the zero-mode is not a good fractional quantum number. 
Instead, in the systems we study, the electronic charge
is conserved and therefore we can show that it is fractionalized by 
properly tallying it.

We find only one normalizable state for each vortex, despite the two
flavors of Dirac fermions in honeycomb lattices. The fact that there is
{\it one} and {\it not two} zero modes is {\it essential} for
fractionalization: doubling the number of zero modes due to two
flavors of Dirac fermions, in addition to the spin degeneracy, would
lead to excitations with the same quantum numbers as ordinary
electrons. We find excitations with charge $Q=e/2$ in the spinless
case, and charge $Q=e$ and spin $S=0$ or charge $Q=0$ and spin $S=1/2$
in the case with spin. Hence, these fractionalized quantum numbers are
similar to those in polyacetelene, but the counting is different in 1D
and 2D, and as we will discuss, there is a $U(1)$ symmetry underlying
the fractionalization in the 2D case which would not lead to
fractionalization in 1D (in which a $\mathbb{Z}^{\ }_{2}$ symmetry is
underlying the effect). 

\begin{figure}
\includegraphics[angle=0,scale=0.4]{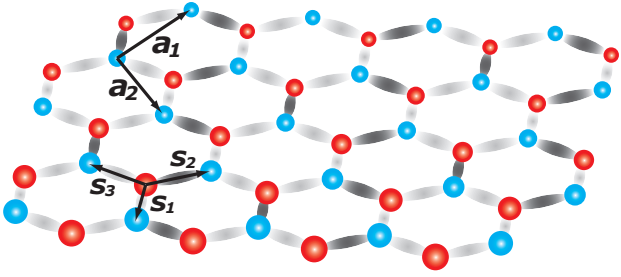}
\caption{
  (color online). The honeycomb lattice $\Lambda$ 
  with lattice spacing $\mathfrak{a}$ and its two triangular
  sub-lattice $\Lambda^{\ }_{A}$ (Red) and $\Lambda^{\ }_{B}$ (Blue). 
  The generators of $\Lambda^{\ }_{A}$ are 
  $\boldsymbol{a}^{\ }_1$ and $\boldsymbol{a}^{\ }_2$. 
  The three vectors $\boldsymbol{s}^{\ }_j$ 
  connect any site from  $\Lambda^{\ }_{A}$ to ist three
  nearest-neighbor sites belonging to  $\Lambda^{\ }_{B}$.
  The Kekul\' e distortion is a modulation of the
  nearest-neighbor hopping amplitude that is indicated by
  representing nearest-neighbor bonds of the honeycomb lattice
  in black (grey) if the hopping amplitude is large (small).
        }
\label{fig:honeycomb}
\end{figure}

The mechanism for fractionalization can be portrayed in a simple form
by considering spinless electrons hopping on a honeycomb lattice with
textured tight-binding hopping amplitudes
and described by the Hamiltonian
\begin{equation}
H=
-\sum_{\boldsymbol{r}\in\Lambda^{\ }_{A}}
\sum_{i=1}^{3}
\left(
t
+ 
\delta t^{\ }_{\boldsymbol{r},i}\right)\;
a^{\dag}_{\boldsymbol{r}}
b^{\   }_{\boldsymbol{r}+\boldsymbol{s}^{\ }_{i}}
+
\mathrm{H.c.}
\label{eq: def non-interacting Hamiltonian}
\end{equation}
Here, $\boldsymbol{s}^{\ }_{i}$ with $i=1,2,3$
connects any site $\boldsymbol{r}$
belonging to the triangular sublattice $\Lambda^{\ }_{A}$
to its three nearest-neighbors
belonging to the triangular sublattice $\Lambda^{\ }_{B}$ 
of the honeycomb lattice as is depicted in Fig.~\ref{fig:honeycomb}.
The fermionic annihilation operators
$a^{\ }_{\boldsymbol{r}}$ and $b^{\ }_{\boldsymbol{r}}$ 
act on $\Lambda^{\ }_{A}$ and $\Lambda^{\ }_{B}$,
respectively, and so do their adjoints. 
Graphene is often described by
Hamiltonian~(\ref{eq: def non-interacting Hamiltonian})
in the single-particle approximation,
neglecting the spin of the electron,
and assuming $\delta t^{\ }_{\boldsymbol{r},i}=0$.
We are going to show that the small variations of the hopping strength, 
$\delta t^{\ }_{\boldsymbol{r},i}$, over the uniform hopping $t$
provide the background on which fractionally charged states can be
constructed. We then discuss how such 
$\delta t^{\ }_{\boldsymbol{r},i}$ can arise from a local order parameter 
that decouples electron-electron interactions.

When $\delta t^{\ }_{\boldsymbol{r},i}=0$,
Hamiltonian~(\ref{eq: def non-interacting Hamiltonian})
can be diagonalized in momentum space,
$
H=
\sum_{\boldsymbol{k}} 
\Phi^{\ }_{\boldsymbol{k}}\,
a^{\dag}_{\boldsymbol{k}} 
b^{\   }_{\boldsymbol{k}} 
+ 
\mathrm{H.c.},
$
$
\Phi^{\ }_{\boldsymbol{k}}= 
-t\sum^3_{j=1} e^{i \boldsymbol{k}\cdot\boldsymbol{s}^{\ }_j}.
$
The single-particle spectrum 
$
\varepsilon^{\ }_{\boldsymbol{k}}=
\pm 
\left|\Phi^{\ }_{\boldsymbol{k}}\right|
$
thus contains two (zero-energy) Dirac points at the zone boundaries 
$
\boldsymbol{K}^{\ }_{\pm}=
\pm\left(\frac{4\pi}{3\sqrt{3}\,\mathfrak{a}},0\right).
$
After linearization,
$
\boldsymbol{k}=\boldsymbol{K}^{\ }_{\pm}+\boldsymbol{p}
$, 
one obtains a spectrum containing two chiral flavors $\pm$, 
$
\mathcal{H}=
\sum_{\boldsymbol{p},\pm}
\phi^{\ }_{\boldsymbol{p}}\,
a^{\dag}_{\boldsymbol{p},\pm} 
b^{\   }_{\boldsymbol{p},\pm} 
+ 
\mathrm{H.c.},
\;
\phi^{\ }_{\boldsymbol{p},\pm}=
\pm v^{\ }_F(p^{\ }_x \pm i p^{\ }_y),
$
with the Dirac conelike structure 
$
\varepsilon^{\ }_{\pm}(\boldsymbol{p})=
\pm v^{\ }_F |\boldsymbol{p}|
$
for the energy dispersion.
(Hereafter, $v^{\ }_F=1$.)

\begin{figure}
\includegraphics[angle=0,scale=0.4]{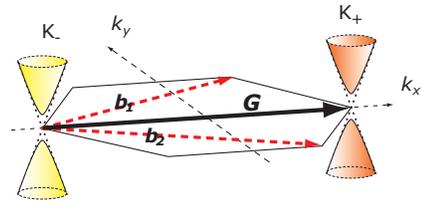}
\caption{
(color online). The first Brillouin zone of the triangular lattice is shown 
 together with the Dirac points $\boldsymbol{K}^{\ }_{\pm}$
 at the zone boundary, the reciprocal vector $\boldsymbol{G}$
 that connects them, and the massive relativistic dispersion
 centered about $\boldsymbol{K}^{\ }_{\pm}$ that opens due
 to a Kekul\'e distortion with $\Delta(\boldsymbol{r})=\Delta^{\ }_{0}$.
        }
\label{fig:Kekule wave vector}
\end{figure}

We shall focus on backgrounds that yield a chiral mixing between the
$\pm$ species. A Kekul\'e texture, depicted in
Fig.~\ref{fig:honeycomb}, provides such a mixing~\cite{Chamon}. The
modulation 
\begin{equation}
\delta t^{\ }_{\boldsymbol{r},i}= 
\Delta(\boldsymbol{r})\;
e^{i \boldsymbol{K}_{+} \cdot \boldsymbol{s}_i}
\, e^{i \boldsymbol{G} \cdot \boldsymbol{r}}/3
+ 
\mathrm{c.c},
\label{eq: def Kekule texture} 
\end{equation}
with wave vector 
$\boldsymbol{G}:=\boldsymbol{K}_{+}-\boldsymbol{K}_{-}$
couples the Dirac points at $\boldsymbol{K}_{\pm}$
as is depicted in Fig.\ \ref{fig:Kekule wave vector}.
Here, we allow for spatial fluctuations 
(on length scales much longer than the lattice spacing $\mathfrak{a}$) 
of the complex-valued order parameter $\Delta(\boldsymbol{r})$.
The phase of $\Delta(\boldsymbol{r})$
controls the ordered hopping texture. To leading order in a gradient expansion,
Hamiltonian~(\ref{eq: def non-interacting Hamiltonian})
subjected to the Kekul\'e texture~(\ref{eq: def Kekule texture})
is given by
$
\mathcal{H}= \int d^{2}\boldsymbol{r} \;
\Psi^{\dag}(\boldsymbol{r}) 
\; \mathcal{K}_{D}(\boldsymbol{r}) \;
\Psi(\boldsymbol{r})
$
with 
$
\Psi^{\dag}(\boldsymbol{r})=
\left(
\begin{array}{cccc}
u^{\dag}_{b}(\boldsymbol{r}) 
&
u^{\dag}_{a}(\boldsymbol{r}) 
&
v^{\dag}_{a}(\boldsymbol{r}) 
&
v^{\dag}_{b}(\boldsymbol{r})
\end{array}
\right)
$
and
\begin{equation}
\mathcal{K}^{\ }_{D}=
\left(\begin{array}{cccc} 
0 
& 
- 
2  
i\partial^{\ }_{{z}}
& 
\Delta(\boldsymbol{r}) 
& 
0 
\\
- 
2 
i\partial^{\ }_{\bar z}  
& 
0 
& 
0 
& 
\Delta(\boldsymbol{r}) 
\\
\bar{\Delta}(\boldsymbol{r}) 
& 
0 
& 
0 
& 
2i\partial^{\ }_{{z}} 
\\
0 
& 
\bar{\Delta}(\boldsymbol{r}) 
& 
2i\partial^{\ }_{\bar z} 
& 
0
\end{array}\right).
\label{eq:Dirac-kernel-real-space}
\end{equation}
We are using the notation
$z=x+iy$, 
$\partial_z=(\partial_x-i\partial_y)/2$, 
with an overline to denote complex conjugation.
Without the Kekul\'e texture~(\ref{eq: def Kekule texture})
leading to $\Delta(\boldsymbol{r})$,
the Dirac kernel~(\ref{eq:Dirac-kernel-real-space}) 
is the single-particle relativistic massless Dirac Hamiltonian 
in 2+1 space-time. 
With the Kekul\'e texture~(\ref{eq: def Kekule texture})
$\Delta(\boldsymbol{r})=\Delta^{\ }_{0}$,
the dispersion takes the simple form 
$
\varepsilon^{\ }_{\pm}(\boldsymbol{p})=
\pm\sqrt{|\boldsymbol{p}|^2+|\Delta^{\ }_{0}|^2}
$, 
i.e., a single-particle mass gap $|\Delta^{\ }_{0}|$ has opened. 
The Dirac kernel~(\ref{eq:Dirac-kernel-real-space}) 
is TRS. TRS originates in the tight-binding hopping elements being real.
Moreover, the transformation law
under
$a^{\ }_{\boldsymbol{r}}\to
-a^{\ }_{\boldsymbol{r}}$
and
$b^{\ }_{\boldsymbol{r}+\boldsymbol{s}^{\ }_{i}}\to
+b^{\ }_{\boldsymbol{r}+\boldsymbol{s}^{\ }_{i}}$
of the single-particle tight-binding%
~(\ref{eq: def non-interacting Hamiltonian})
ensures that any positive energy eigenstate 
of the Dirac kernel~(\ref{eq:Dirac-kernel-real-space}) 
can be matched to a negative energy eigenstate, 
while only zero-modes can be left unmatched.
We shall call this property sublattice symmetry (SLS)
\cite{note1}.

The order parameter $\Delta(\boldsymbol{r})=\Delta^{\ }_{0}$ 
can be complex valued, but he spectral mass gap is real. 
This suggests that the phase of $\Delta^{\ }_{0}$
is redundant. In fact, it can be removed 
from Eq.~(\ref{eq:Dirac-kernel-real-space}) 
with a chiral transformation that rotates the phases
of the $\pm$ species by opposite angles. 
This is not true anymore if
the phase of the order parameter $\Delta(\boldsymbol{r})$ 
varies in space, and, in particular, if it contains vortices. 
The latter situation leads to fractionalization.

We assume that
\begin{equation}
\Delta(\boldsymbol{r})= 
\Delta^{\ }_{0}(r)\,e^{i(\alpha+n\theta)}
\label{eq: charge n vortex}
\end{equation}
where $\Delta^{\ }_{0}(r)>0$,
$n\in\mathbb{Z}$ for a single-valued order parameter,
and we are using the polar coordinates $z=r\exp(i\theta)$.
We are seeking eigenstates of the Dirac kernel%
~(\ref{eq:Dirac-kernel-real-space}) 
with vanishing energy that are normalizable, i.e.,
\begin{subequations}
\label{eq:Dirac-Kernel-zero-mode-A}
\begin{eqnarray}
\label{eq:Dirac-Kernel-zro-mode-A-1} 
\left( 
\partial^{\ }_{r} 
- 
ir^{-1}\partial^{\ }_{\theta})\; u^{\ }_{a}(\boldsymbol{r})
+ 
ie^{i\theta}\Delta(\boldsymbol{r}
\right)\; 
v^{\ }_{a}(\boldsymbol{r})&=&0,
\\
\label{eq:Dirac-Kernel-zro-mode-A-2}
ie^{-i\theta}\bar{\Delta}(\boldsymbol{r})\; 
u^{\ }_{a}(\boldsymbol{r})
-
\left(
\partial_{r}
+
ir^{-1}\partial^{\ }_{\theta})\; v^{\ }_{a}(\boldsymbol{r}
\right)&=&0,
\end{eqnarray}
\end{subequations}
holds on sublattice $\Lambda^{\ }_{A}$ 
while two more equations obtained from
Eq.~(\ref{eq:Dirac-Kernel-zero-mode-A})
with the substitutions
$u^{\ }_{a}\to  u^{\ }_{b}$,
$v^{\ }_{a}\to  v^{\ }_{b}$,
and
$\theta\to-\theta$ 
must also hold  on sublattice $\Lambda^{\ }_{B}$.
The same equations were studied in Refs.~\cite{Read-Green,Jackiw-Rossi}
from a different perspective since the textured gap in these works
is associated to a superconducting order parameter.
Hence, electric charge is not a conserved quantum number in 
Refs.~\cite{Read-Green,Jackiw-Rossi}. 
There are $|n|$ independent normalizable zero modes,
which are either supported on sublattice $\Lambda_A$ when $n \le -1$
or on sublattice $\Lambda_B$ when $n\ge1$.

We assume that $n=-1$
in which case the single-valued and normalizable wave
functions for the zero mode is
\begin{equation}
\begin{split}
u^{\ }_{a}(r,\theta)=& 
\frac{
e^{i\left(\frac{\alpha}{2}+\frac{\pi}{4}\right)}
     }
     {2
\sqrt{\pi}
     }\, 
\;
\frac{
e^{-\int_0^r dr'\,\Delta_0(r')}
     }
     {
\sqrt{\int_0^\infty dr\,r\,e^{-2\int_0^r dr'\,\Delta^{\ }_0(r')}}
     },
\\
v^{\ }_{a}(r,\theta)=&
\bar{u}^{\ }_{a}(r,\theta).
\end{split} 
\label{eq: wave fct zero mode}
\end{equation}
Its support is on sublattice $\Lambda^{\ }_A$.
The zero mode when $n=1$ is obtained from
Eq.~(\ref{eq: wave fct zero mode})
with the substitutions
$u^{\ }_{a}(r,\theta)\to v^{\ }_{b}(r,\theta)$
and
$v^{\ }_{a}(r,\theta)\to u^{\ }_{b}(r,\theta)$.
When $n=1$, the support of the zero mode 
is on sublattice $\Lambda^{\ }_B$. 
The wave function~(\ref{eq: wave fct zero mode})
decays exponentially fast away from the core
of the vortex~(\ref{eq: charge n vortex})
at the origin of the complex plane.
Its localization length is set by
the gap value $\Delta^{\ }_0$ reached when
$\Delta^{\ }_0(r)$ saturates sufficiently far from the origin.
If the Kekul\' e texture~(\ref{eq: def Kekule texture})
supports a pair of vortices a large distance $R$ apart,
then the Dirac kernel~(\ref{eq:Dirac-kernel-real-space}) 
has two eigenstates whose energy eigenvalues are exponentially small.

To obtain the charge bound to a vortex, one has to study 
$\delta\nu(\boldsymbol{r},\varepsilon)\equiv
\nu^{\ }_{|n|=1}(\boldsymbol{r},\varepsilon)
-
\nu^{\ }_{|n|=0}(\boldsymbol{r},\varepsilon)
$ 
where $\nu^{\ }_{n}(\boldsymbol{r},\varepsilon)$
is the local density of states (LDOS)
of the Dirac kernel~(\ref{eq:Dirac-kernel-real-space})
in the presence of
the mass twist~(\ref{eq: charge n vortex}). 
Because of the SLS,
to any negative eigenstate of the Dirac kernel, 
$\psi^{\ }_{-\varepsilon}(\boldsymbol{r})$,
there corresponds a positive energy state, 
$\psi^{\ }_{+\varepsilon}(\boldsymbol{r})$,
related to $\psi^{\ }_{-\varepsilon}(\boldsymbol{r})$
by a unitary transformation.
Hence, the LDOS
$\nu(\boldsymbol{r},\varepsilon)\equiv
\sum_{\varepsilon'}
\psi^{\dag}_{\varepsilon'}(\boldsymbol{r})
\psi^{\   }_{\varepsilon'}(\boldsymbol{r})
\delta(\varepsilon-\varepsilon')$ 
is symmetric with respect to zero energy
and negative and positive energy eigenstates
contribute equally to the local density
$
\nu(\boldsymbol{r})\equiv
\lim_{\varepsilon\to\infty}
\int_{-\varepsilon}^{+\varepsilon}d\varepsilon'
\nu(\boldsymbol{r},\varepsilon')
$.
In the presence of the zero mode $\psi^{\ }_0(\boldsymbol{r})$
this spectral symmetry 
together with the conservation of the total number of states imply
\begin{equation}
\int d^2 \boldsymbol{r} 
\left( 
2\int_{-\infty}^{0^{-}} 
\delta\nu(\boldsymbol{r},\varepsilon) \,d\varepsilon 
+ 
|\psi^{\ }_0(\boldsymbol{r})|^2 
\right)=0.
\label{eq:Density-state-balance I}
\end{equation}
But the \textit{single} zero mode $\psi^{\ }_0(\boldsymbol{r})$ 
is normalized to one and
\begin{equation}
\int d^2 \boldsymbol{r} 
\int_{-\infty}^{0^{-}} 
\delta\nu(\boldsymbol{r},\varepsilon) \,d\varepsilon=
-1/2.
\label{eq:Density-state-valence II}
\end{equation}
According to Eq.~(\ref{eq:Density-state-valence II}), 
the valence band has a deficit of half of a state, 
as does the conduction band. Thus the difference in net charge
between a fully occupied valence band with and without the vortex is
$-e/2$. The total charge of a closed system must still be an integer.
This can be understood as follows. Vortices must be present in
pairs (for example, if a finite system is to satisfy
periodic boundary conditions or for energetic reasons
in the thermodynamic limit), and while the charge around each
vortex is half that of an electron, the fact that the
vortices appear in pairs ensures that the total charge is an
integer multiple of $e$. The mechanism for fractionalization is
similar to that in 1D~\cite{Su1980}, but there the fractional charge
stands in domain walls between two degenerate vacua, which must
appear in pairs (again if the system is to satisfy periodic boundary
conditions). In 1D, however, one needs a discrete symmetry to
fractionalized charge. For example, for charge $e/2$ states one needs a
$\mathbb{Z}^{\ }_{2}$ symmetry, which can be understood generically in terms
of an accumulation of charge along phase twists that have a natural
interpretation within a bosonization scheme~\cite{Goldstone1981}. These
arguments do not extend to 2D, which is our concern in this letter, 
as there is no simple connection between phase twist and charge
as in the 1D bosonization scheme. The fact that the fractional states
must occur in pair, nonetheless, follows from the common requirement
for the $U(1)$ mass vortices in 2D and the $\mathbb{Z}^{\ }_2$ domain
walls that these defects occur in pairs to satisfy appropriate
boundary conditions.

In the case of electrons with spin, two independent zero modes that carry
the spin quantum number are expected. Hence, there is,
in total, a full electronic state missing in the valence band for each
vortex. Consequently, unoccupied, singly, and doubly occupied zero
energy states correspond to the charge and spin quantum numbers 
$(Q=-e, S=0)$, $(Q=0, S=1/2)$, and $(Q=+e, S=0)$, respectively. 
Notice that, had we obtained \textit{two} 
normalizable solutions for a vortex or an
anti-vortex as opposed of \textit{one} as in 
Eq.~(\ref{eq: wave fct zero mode}), 
all the quantum numbers would be ``doubled'' and simply coincide with
those of ordinary electrons.

These arguments for fractionalization require existence of the
mid-gap state and the SLS of the spectrum. 
SLS is broken when a next-nearest
neighbor hopping $t'$ is introduced.
A $t'$ brings an overall shift and
a quadratic term into the diagonal sector of the Hamiltonian, $\delta
\mathcal{H}'= (3t'+\frac{9 t'\mathfrak{a}^2}{4} p^2)\;\openone$, 
up to second order in $\boldsymbol{p}$. 
This perturbation alters the energy spectrum, 
but Dirac points (in the absence of the background) remain when the
systems is at half-filling. The textured background still opens a gap
at the Dirac points and there is a mid-gap state if there is a mass twist. 
However, the mid-gap state is not exactly at the center between the valence 
and conduction bands, being shifted by 
$\delta \varepsilon^{(1)}=\frac{t'\Delta_0}{t^2} \Delta^{\ }_0$ 
to first order $t'$. 
As long as this shift is small compared to the gap, 
the mid-gap solution is robust against breaking of the SLS.
The fractionalized quantum number persists along with the
single bound state, but the argument for charge $e/2$ is not
valid. Starting from zero $t'$, turning it adiabatically, one can
argue that fractionalization should remain but take irrational
values as in 1D~\cite{Rice1982,Jackiw-Semenoff,Kivelson1983}.

We now turn to a mechanism for generating spontaneously the Kekul\'e
distortions from repulsive electron-electron interactions. 
We assume the nearest-neighbor interaction
$
H':= 
-V 
\sum_{\boldsymbol{r}\in\Lambda^{\ }_A} 
\sum_{j=1}^{3}  
a^{\dag}_{\boldsymbol{r}}\, 
b^{\   }_{\boldsymbol{r}+\boldsymbol{s}^{\ }_j}\,
b^{\dag}_{\boldsymbol{r}+\boldsymbol{s}^{\ }_j}\,
a^{\   }_{\boldsymbol{r}}
$
with $V$ the interaction strength.
Evidently, this interaction preserves TRS,
the point group symmetry of the honeycomb lattice,
and SLS (up to the total number operator).
After linearization around the unperturbed Dirac points,
a mean-field decoupling of this interaction with respect to the
order parameter
\begin{equation}
-\frac{2} {3V}
\Delta=
\langle 
b^{\dag}_{-}(\boldsymbol{r}+\boldsymbol{s}_j) \; 
a^{    }_{+}(\boldsymbol{r}) 
\rangle 
= 
\langle 
a^{\dag}_{-}(\boldsymbol{r}) \;
b^{\   }_{+}(\boldsymbol{r}+\boldsymbol{s}_j) 
\rangle
\label{eq: Kekule mean-field channel}
\end{equation}
coincides with the case of the Kekul\'e distortions. 
If the system is at half-filling and zero temperature,
all negative energy states are occupied and the self-consistent
equation
$
1= 
\frac{3V}{2 |\Lambda^{\ }_A|} \sum_{\boldsymbol{p}} 
\frac{
\exp
\left(
i\boldsymbol{p}\cdot\boldsymbol{s}^{\ }_j
\right)
     }
     {
\sqrt{
|\boldsymbol{p}|^2 
+
|\Delta|^2}
     }
$
follows. The sum can be replaced by an integral up to an
appropriate momentum cutoff, 
$\Lambda\mathfrak{a}\sim\frac{2\pi^{1/2}}{3^{3/4}}$, 
here chosen so as to match
the total number of states in the effective theory to that in the 
microscopic theory. 
As this integral converges even without the $|\Delta|^2$ term, 
neglecting it yields
$
1\leq
\frac{3\sqrt{3}V\Lambda\mathfrak{a}}{8|t|} 
[J^{\ }_{0}(\Lambda\mathfrak{a})\;{H}^{\ }_{-1}(\Lambda\mathfrak{a})
-
J^{\ }_{-1}(\Lambda\mathfrak{a})\;{H}^{\ }_{0}(\Lambda\mathfrak{a})].
$
Here, $J^{\ }_{\nu}(z)$ is a Bessel function and
${H}^{\ }_{\nu}(z)$ is a Struve function.
Insertion of the numerical value for $\Lambda\mathfrak{a}$ 
then yields the lower bound
$
V^{\ }_c\geq
\frac{|t|}{0.52465}=1.906 |t|.
$
Self-consistency demands a repulsive interaction. 
(The very same mean-field theory also renormalizes
the uniform hopping: $t\to t+\delta t$, $\delta t\sim0.1 V$.)
We note that the condition for the Kekul\'e pattern to form could
potentially be engineered in a gas of dipolar fermionic atoms trapped
in an optical honeycomb lattice, where the ratio $V/t$ could be tuned.

At last, we need a vortex on top of the Kekul\'e distortion%
~(\ref{eq: def Kekule texture})
for fractionalization to happen.
Here, we face a difficulty in that the chiral symmetry
of Eq.~(\ref{eq:Dirac-kernel-real-space}) is an artifact of
the linearization. Integration of the fermions on
the lattice yields an effective action 
for the phase of $\Delta$ 
with a discrete symmetry $\mathbb{Z}^{\ }_{3}$. 
Vortices are then confined at sufficiently long distances due to
a pinning potential for the phase of $\Delta$. Fractionalization
can only be observed on length scales much larger than $\Delta^{\ }_{0}$
yet much smaller than the confining length scale, a function
of $\Delta^{\ }_{0}/t$. This difficulty can be overcome by a
small breaking of the point-group symmetry of the honeycomb lattice
by assuming anisotropic hopping between nearest-neighbor sites.
By continuity, the only effect of such a small anisotropy
is to move the Dirac points away from the boundaries of the 
first Brillouin Zone, as is confirmed by an exact computation
of the spectrum in Ref.~\onlinecite{Hasegawa06}. 
The Kekul\'e wave vector 
$\boldsymbol{G}$ in Eq.~(\ref{eq: def Kekule texture}) 
is then not anymore commensurate to the reciprocal lattice, 
integration over the fermions does not produce a pinning
potential for the phase of Kekul\'e modulation, and vortices
can proliferate as a result of temperature-induced fluctuations
(their bare logarithmic interaction is screened for temperatures 
above the Kosterlitz-Thouless transition temperature).

In summary, we presented a mechanism in graphenelike 2D condensed
matter systems that realizes charge fractionalization with TRS.  The
mechanism is based on an effective low-energy Hamiltonian of the Dirac
type with a textured mass with phase twists due to vortices.  When the
vortices are far apart, each vortex carry a zero-mode responsible for
local charge fractionalization.  This mechanism is a condensed matter
2+1 space-time analogue to 3+1 space-time cosmic strings.


We would like to thank Eduardo Fradkin for illuminating comments
about the confinement of vortices for a commensurate
Kekul\'e modulation.
This work is supported by the NSF grant DMR-0305482 (C.-Y.~H and
C.~C.).

\vspace{-.2in}


\bibliography{refer.bib}

\begin{thebibliography}{23}
\bibitem[{not()}]{note1}
\bibinfo{note}{We reserve the terminology of particle-hole symmetry to
  single-particle Hamiltonians of the Bogoliubov-de-Gennes type such as in
  Refs.\ \onlinecite{Read-Green} and \onlinecite{Jackiw-Rossi}.}
\end{thebibliography}


\end{document}